\documentclass[12pt]{article}
\usepackage{graphicx}
\usepackage{bm}

\newcommand{\eq}[1]{Eq.~(\ref{#1})}
\newcommand{\eqs}[2]{Eqs.~(\ref{#1}) and (\ref{#2})}
\newcommand{\dpar}[2]{\frac{\partial#1}{\partial#2}}
\def\be{\begin{equation}}
\def\ee{\end{equation}}
\def\ba{\begin{eqnarray}}
\def\ea{\end{eqnarray}}
\def\half{{1 \over 2}}
\def\part{\partial}

\def\eps{\epsilon}

\def\Ome{\Omega}
\def\lam{\lambda}

\def\Gam{\Gamma}
\def\D{\Delta}

\def\cA{{\cal A}}

\def\tA{\widetilde{A}}
\def\bPsi{\bar{\Psi}}
\def\rtip{r_{tip}}
\def\phirad{\phi_{rad}}
\def\Sent{S_{ent}}

\begin{document}
\title{\Large {\bf Entanglement entropy of a superflow}}
\author{\normalsize Sergei Khlebnikov and Akhil Sheoran \\ 
\normalsize{\it Department of Physics and Astronomy, Purdue University,} \\
\normalsize{\it West Lafayette, IN 47907, USA}
}
\date{}

\maketitle

\begin{abstract}
We consider the theory of
$N$ free Dirac fermions with a uniformly winding mass, $m e^{iqx}$, in two spacetime
dimensions. This theory (which describes for instance a superconducting current 
in an $N$-channel wire) has been proposed to have a higher-spin gravity with scalar matter
as the large-$N$ dual. To order $m^2$, however, thermodynamic quantities in it can be 
computed using standard general relativity instead. Here, we consider the
question if the same is true for the entanglement entropy (EE). 
By comparing results obtained on two sides of the duality, we find that general relativity
indeed accounts correctly for the EE of an interval to order $m^2$ (and all orders in $q$).

\end{abstract}

\section{Introduction} \label{sec:intro}
In semiclassical gravity, the Gibbons-Hawking (GH) formula \cite{Gibbons&Hawking} 
interprets the Euclidean effective action of a spacetime as the 
free energy (of gravity plus matter) and thus provides a method of computation of
gravitational entropy. If the spacetime is asymptotically anti-de Sitter (AdS), 
the same effective action also determines the entropy of the dual conformal field theory 
(CFT) in the context of AdS/CFT duality \cite{Maldacena}. Furthermore,
if the CFT is deformed by a relevant operator, the 
correspondence \cite{Gubser&al,Witten:1998} between operators in the CFT and fields in AdS
tells us the precise way in which the entropy of the CFT can be computed from the
action of a deformed spacetime. 

Replacing the thermal density matrix in the von Neumann entropy formula with one obtained
by integrating out a subset of degrees of freedom defines entanglement entropy (EE).
It has been shown
\cite{Casini&al:Derivation} that the EE of a spherical region in a CFT vacuum is the same as
the entropy of a thermal state in an auxiliary hyperbolic space and that
application of the AdS/CFT correspondence to this thermal state, 
in the case when the gravitational dual is standard general relativity, reproduces the result
of the minimal-surface formula of Ryu and 
Takayanagi \cite{Ryu&Takayanagi}. Furthermore, it
has been argued \cite{Lewkowycz&Maldacena}, in a similar context,
that the EE can be related to the area of a minimal surface
directly, without going through a thermal state as an intermediary; that argument applies 
to a region of any shape.

In studies of the AdS/CFT correspondence, of special interest 
are cases when the CFT is solvable, e.g., a free theory. One such case is the theory
of $N$ free Dirac fermions
in two spacetime dimensions ($d=2$), which has been conjectured \cite{Gaberdiel&Gopakumar}
to be dual, in the large $N$ limit, to a higher-spin gravity with scalar matter in AdS$_3$ 
\cite{Prokushkin&Vasiliev}. Deformation of the CFT by a mass term corresponds to 
a nontrivial profile of a scalar. A scalar of amplitude $A$ sources the 
higher-spin fields at the same (quadratic \cite{Prokushkin&Vasiliev}) order 
in $A$ as it sources the metric correction. To this order, however, neither those 
fields nor the metric correction contribute to variation of the action; the latter 
is given simply by the action of the scalar on the undeformed background. 
As a result, the $O(A^2)$ correction to the thermal entropy can be computed from the GH formula
as if the dual theory were standard general relativity. 

One may then wonder if to 
order $A^2$ the generalized GH formula \cite{Lewkowycz&Maldacena} 
also works the same way as 
in general relativity, in particular, if the Ryu-Takayanagi (RT) formula for the EE still 
applies.
In this paper, we address this question by making comparisons between results obtained on two
sides of the duality.

The theory of $N$ Dirac fermions in $d=2$ is of interest and perhaps some practical
importance in its own right. It can be used 
to describe a quantum wire with $N$ transverse channels. If we identify the chiral charge
of the fermion with the electric charge in the wire, the fermion mass corresponds to
a superconducting pairing amplitude induced for instance by proximity to a larger 
superconductor. 
The mass winding along the wire as $\tilde{m}(x) = m e^{iqx}$, where $m > 0$ and $q$ are real
constants, corresponds to a state with non-zero supercurrent. 
According to the preceding, we may hope to use standard general relativity to compute 
both the thermal and entanglement entropies to order $m^2$ and all orders in $q$.

Our main focus will be a non-vacuum, thermal state in the CFT, related
holographically to the BTZ black hole \cite{BTZ}. In this case, there is potentially a small
parameter $m/T$, where $T$ is the temperature. On the gravity side, a small $m/T$ corresponds
to a small deformation of the BTZ background. In contrast, if $m$ is 
larger than both $T$ and $|q|$, the deformation cannot be small everywhere, and the only small
dimensionless parameter proportional to $m$ is $m l$, where $l$ is the length of
the entangling region. We briefly discuss this case towards the end of the paper.

We start in Sec.~\ref{sec:asymp} with establishing a relation between the scalar amplitude
$A$ in gravity and the fermion mass $m$ in the CFT. The relation uses only
the leading near-boundary
asymptotics of the scalar and amounts to a computation of the scalar two-point function in 
the special, logarithmic case of the AdS/CFT 
correspondence. 
In Sec.~\ref{sec:therm}, we consider the thermal entropy 
and in Sec.~\ref{sec:entan} the EE of an interval of length $l$ in a thermal state of the CFT. 
In both cases, we find agreement between $O(m^2)$ results obtained on two sides of 
the duality. For the EE, our evidence in mostly numerical, but in the case of 
a short interval, which we consider in Sec.~\ref{sec:short}, some results can be extracted 
analytically. In particular, the coefficient of the 
leading $(ml)^2 \ln^2 l$ term in the $O(m^2)$ correction obtained from the RT formula 
is found to coincide with that obtained in Ref.~\cite{Casini&al} directly in field theory. 
We present a discussion of our results in Sec.~\ref{sec:disc}.

\section{Asymptotic analysis} 
\label{sec:asymp}
We consider deformations of a Euclidean non-rotating BTZ black hole \cite{BTZ}, 
due to a small-amplitude 
complex scalar field $\phi$. The metric is of the form
\be
ds^2 = r^2 d\tau^2 + F(r) dx^2 + G(r) dr^2 \, .
\label{ds2}
\ee
The asymptotic (near-boundary) region corresponds to large $r$, where 
$F(r) \to r^2$ and $G(r) \to 1 / r^2$, after we set the AdS radius to 
unity. The coordinates $x$ and $\tau$ are subject to the following peridocities:
\ba
x & \sim & x + L_x \, , \label{perx} \\
\tau & \sim & \tau + \beta \, . \label{pert}
\ea
The GH formula interprets $T = 1/\beta$ as the temperature of the spacetime. According
to the AdS/CFT correspondence \cite{Maldacena}, it is also the temperature in the dual CFT,
while $L_x$ is the length of the spatial circle on which the CFT lives.
 
For the undeformed black hole 
\be
F(r) = F_0(r) \equiv r^2 + \alpha^2 \, , \hspace{3em} G(r) = G_0(r) \equiv [F_0(r)]^{-1} \, ,
\label{btz}
\ee
with
\be
\alpha = 2\pi T \, .
\label{alpha}
\ee
The entropy of this space is $S_0 = \alpha L_x / (4 G_N)$ \cite{BTZ},
where $G_N$ is Newton's constant. We now proceed to finding corrections to this result, 
due to the presence of the scalar.

The Gibbons-Hawking method of finding gravitational entropy
requires computing the Euclidean action including, in our case, both gravity and the
scalar. If we only want the action to the leading ($A^2$) order in the scalar amplitude $A$, 
we do not need to consider changes to the gravitational part of the action, since that part 
was extremal for $A = 0$. This leaves us with the bilinear action of the scalar,
\be
I_E = \int d^3 x \sqrt{g} \left( g^{mn} \partial_m \phi^* \partial_n \phi 
+ M^2 \phi^* \phi \right) \, ,
\label{IE}
\ee
where $g$ refers to the metric of the undeformed background.
On the equations of motion, this reduces to the boundary term
\be
I_E = \int d\tau dx \left. (r g^{rr} \phi^* \partial_r \phi) \right|_{r=r_m} \, ,
\label{Ibou}
\ee
where $r_m$ is some large value of the radius. 

Here, we focus on the logarithmic case $M^2 = -1$, corresponding, via the standard
AdS/CFT dictionary \cite{Gubser&al,Witten:1998}, to an operator of dimension $\D = 1$ 
in the CFT. We restrict attention to static $x$-dependent solutions of the form
\be
\phi(x,r) = e^{iqx} \phirad(r) \, ,
\label{phi}
\ee
with a constant momentum $q$ in the $x$ direction. 
The leading and first subleading terms in the asymptotic of the scalar in this case can be
combined into
\be
\phirad(r) = \frac{A}{r} \ln (r \xi) + \dots  \, ,
\label{phiasymp}
\ee
where $A$ and $\xi$ are constants, and the dots stand for terms suppressed by inverse 
powers of $r$. The constant $\xi$ is a function of $q$ and $T$,
\be
\xi = \xi(q, T) \, ,
\label{xidep}
\ee
but, in the present case, not of $A$ itself. 
For $q = 0$, $\xi$ can be considered as a holographic definition of the correlation 
length. We will occasionally use this terminology also for $q \neq 0$.

For a scalar described by a bilinear action on the undeformed BTZ background, the solution
to the equations of motion is readily available, 
and one can read off it the full dependence of 
$\xi$ on its arguments. We will make use of this solution in the next section. Here, let us
observe that both \eqs{Ibou}{phiasymp} depend only on the asymptotic form of the metric
and so are a bit more general than the case at hand. 
We can use them to obtain a general relation of 
$A$ to the mass parameter of the CFT. 
It will apply for instance also in the case of vacuum AdS, which
has the same large-$r$ asymptotic as the BTZ metric (\ref{btz})
but a different topology.

Substituting \eq{phiasymp} in (\ref{Ibou}), we find that the action
is divergent in the limit $r_m \to \infty$ and so needs to be renormalized. This is achieved
by adding a boundary counterterm \cite{Klebanov&Witten}. 
We do that by first stepping away from the logarithmic case,
i.e., considering $M^2$ somewhat above $-1$, so that
\be
\lam \equiv 1 + (1 + M^2)^{1/2} = 1 + s
\label{lam}
\ee
with $0< s < 1$, and then taking the limit $s\to 0$. The counterpart of \eq{phiasymp} is
\be
\phirad(r) = A_s r^{s-1} \left[ 1 - (\xi r)^{-2s} \right] + \dots \, ,
\label{phis}
\ee
and the renormalized action, computed by the method of Ref.~\cite{Klebanov&Witten}, is
\be
I_{E,ren} = 2 s \int d\tau dx |A_s|^2 \xi^{-2s} \, .
\label{Iren}
\ee
For \eq{phis} to reproduce (\ref{phiasymp}) in the limit $s\to 0$, 
$A_s$ must go infinity as $A/ (2s)$ with a constant $A$. 
We then see that the logarithmic case
requires an additional subtraction, of the $O(1/s)$ term in (\ref{Iren}).
After this subtraction, the renormalized action for $s = 0$ becomes
\be
I_{E,ren} = - \int d\tau dx |A|^2 \ln(\mu  \xi) \, ,
\label{Iren2}
\ee
where $\mu$ is a normalization momentum.

By the standard dictionary \cite{Gubser&al,Witten:1998}, $I_{E,ren}$ generates the two-point
function of an operator of  
dimension $\D = 1$ in the dual CFT. The result can be compared to that obtained directly in
in the field theory of a $N$-component Dirac fermion in $d=2$. The operator
in question there is $O = \bPsi_R \Psi_L$, where $\Psi_{L,R}$ are the left- and 
right-moving components of the fermion field. The source of $O$ in this
theory is the position-dependent fermion mass $m e^{iqx}$, where we can choose $m$ to be real 
and positive. The two-point function, computed from 
a one-loop diagram is, to logarithmic accuracy,
$(N/2\pi) \ln (\mu / Q)$ where $Q$ is the largest of the three mass scales: $m,|q|$, and $T$.
Comparing this to \eq{Iren2}, 
we see that, if the results on two sides of the duality are to match, 
the amplitude $A$ in the gravitational theory must be related to $m$ as follows:
\be
A = m \sqrt{\frac{N}{2\pi}} \, ,
\label{A}
\ee
up to an inessential constant phase factor. 

We conclude this section with a comment concerning 
the relative magnitude of $q$ and $m$. In a superconductor,
$m$ represents the energy gap and $q$ the superflow momentum per Cooper pair.  
It is well known that, in a conventional intrinsic superconductor, the $q / m$ ratio 
cannot be arbitrarily large: at a sufficiently large $q$, the gap goes to zero signaling
transition to the normal state \cite{Tinkham}. 
Here, however, we consider $q$ and $m$ as independent 
parameters, as may be the case when superconductivity is induced by the proximity effect.
Accordingly, we extend our analysis to arbitrarily large $q$ (and indeed
will not detect any obstacles to doing so).

\section{Thermal entropy}
\label{sec:therm}
The solution for the radial part of the scalar on the rigid BTZ background (\ref{btz}) 
is
\be
\phirad(r) = \tA (r^2 + \alpha^2)^{-\lam/2}
{}_2 F_1\left( a, b, 1;  \frac{r^2}{r^2 + \alpha^2} \right) 
\label{phirad}
\ee
where $\tA$ is a constant amplitude, $\lam$ is related to the mass squared of the scalar 
by \eq{lam}, and ${}_2 F_1$ is the hypergeometric function with 
\be
a = \half \left( \lambda + \frac{i q}{\alpha} \right) \, , \hspace{3em} 
b = \half \left( \lambda - \frac{i q}{\alpha} \right) \, .
\label{ab}
\ee
As before, we focus on the case $\lam =1$ ($M^2 = - 1$), when the asymptotic of
${}_2 F_1$ is logarithmic \cite{AS:Hyper}:
\be
F(a, b, a + b; z) = \frac{\Gam(a + b)}{\Gam(a) \Gam(b)} 
\left[ 2 \psi(1) - \psi(a) - \psi(b) - \ln(1 - z) \right] + O[(1-z)\ln(1-z)] \, ,
\label{F}
\ee
where $\Gam(\cdot)$ and $\psi(\cdot)$ are the gamma and digamma functions, respectively.
The asymptotic form of the solution then agrees with \eq{phiasymp}, with
\ba
A & = & \frac{2 \tA}{\pi} \cosh \frac{\pi q}{2\alpha} \, , \label{tA} \\
\ln \xi & = & - \ln \alpha + \psi(1) 
- \mbox{Re~} \psi\left( \half + \frac{i q}{2\alpha} \right) \, . \label{lnxi}
\ea
The real part of the digamma function in (\ref{lnxi}) is a monotonically increasing 
function of $q^2$, approaching $\ln (q / \alpha)$ in the limit of large $q / \alpha$.
Note, then, that in this limit $\xi$ becomes independent of the temperature, and
to logarithmic accuracy $\ln \xi = - \ln q$. 

The small-amplitude condition for the scalar, under which the metric deformation is small and 
\eq{phirad} is applicable, is 
\be
\kappa A \ll \xi^{-1} \, ,
\label{cond_A}
\ee
where $\kappa^2 = 8 \pi G_N$, and $G_N$ is Newton's constant. Written in terms of the fermion 
mass in the dual CFT, this becomes
\be
m \ll \xi^{-1}  \, ,
\label{cond_m}
\ee
where we have used the relation $G_N = 3/ (2 N)$ \cite{Brown&Henneaux} and \eq{A}. 
Since $\xi$ is determined by the larger of $\alpha = 2\pi T$ and $q$, the condition
(\ref{cond_m}) is automatically satisfied if $m \ll 2\pi T$. 
This is analogous to the Ginzburg-Landau limit in a superconductor.

In accordance with the GH formula, the action (\ref{Iren}) with the $\tau$ integral removed
is interpreted as a correction to the free energy of the black hole:
\be
\delta \Ome_{grav} = -L_x |A|^2 \ln (\mu \xi) \, ,
\label{Ome_grav}
\ee
where we now have an explicit expression for $\ln \xi$, \eq{lnxi}. In view of \eq{A}, this
can also be written as
\be
\left. \dpar{\Ome_{grav}}{m^2} \right|_{m=0} = \frac{N L_x}{2\pi} \ln (\mu \xi) \, .
\label{dgrav}
\ee
The corresponding correction to the entropy then follows from the thermodynamics formula
$S_{grav} = - \part \Ome_{grav} / \part T$.
Note that dependence on the normalization point $\mu$ disappears upon taking the derivative
with respect to $T$.

We now compare this result to the free energy of a multiplet of $N$ free Dirac fermions 
in two spacetime dimensions. The mass Lagrangian density (in the Lorentzian signature) is
\be
L_m = - m \left( e^{-iqx} \bPsi_L \Psi_R + e^{iqx} \bPsi_R \Psi_L \right) \, ,
\label{Lm}
\ee
where $\Psi_{L,R}$ are the left- and right-moving component of the multiplets
and $m$ is a positive constant. A chiral transformation will remove the factor $e^{iqx}$ 
in \eq{Lm}, at the price of an additional term appearing in the full Lagrangian:
\be
L = L' - \frac{q}{2} \left( \Psi_R^\dagger \Psi_R + \Psi_L^\dagger \Psi_L \right) \, ,
\label{Ladd}
\ee
where $L'$ is the Dirac Lagrangian density with a constant mass $m$.
In a superconductor, the quantity in the brackets in (\ref{Ladd}) 
represents the electric current:
recall that we identify the electric charge with the chiral charge of the Dirac fermion 
and set $v_F = 1$. 

The latter form of the Lagrangian is convenient for finding the spectrum of elementary
excitations. There are two branches, with energies
\be
\eps_\pm(k) = \sqrt{k^2 + m^2} \pm \half q \equiv \eps_0(k) \pm \half q \, .
\label{eps}
\ee
The free energy $\Ome_F$ of these fermions is that of the grand canonical ensemble
with the chemical potential set to zero.
Corrections to $\Ome_F$ due to the fermion mass are contained in 
\be
\dpar{\Ome_F}{m^2} = N \sum_k \frac{1}{2\eps_0(k)} \left[ n_F(\eps_+) + n_F(\eps_-) \right] \, ,
\label{ddm2}
\ee
where $n_F(\eps) =(e^{\beta \eps} + 1)^{-1}$ is the Fermi distribution.
Our use of the grand canonical ensemble for an isolated system implies that we are working
in the thermodynamic limit, when the sum over $k$ can be replaced with an integral.
A convenient way to compare the result to the one obtained on the gravity side is to 
expand both expressions in powers of $q^2$ and compare them term by term. The constant, 
power zero, term in (\ref{ddm2}) is obtained by replacing $\eps_\pm$ with $\eps_0$ and is
logarithmic:
\be
\frac{N L_x}{2\pi} \int \frac{dk}{\eps_0(k)} n_F(\eps_0)  
= \frac{N L_x}{2\pi} \left( \ln \frac{T}{m} + \mbox{const} \right) .
\label{logT}
\ee
This differs from the corresponding term in (\ref{dgrav}) by a $T$-independent constant.
So, the results for the entropy match. Nonzero powers of $q^2$, obtained by expanding 
Eq. (\ref{ddm2}) in $q$, have finite limits at $m=0$. These limits are seen to coincide, term
by term, with powers of $q^2$ obtained by expanding $\ln \xi$ in \eq{dgrav}.

\section{Entanglement entropy}
\label{sec:entan}
The Ryu-Takayanagi (RT) formula \cite{Ryu&Takayanagi}
for the EE  has been argued \cite{Lewkowycz&Maldacena} 
to follow from a generalized GH formula, combined with a holographic interpretation
of the gravitational partition function.  
In general, one should not expect the RT formula to apply to
higher-spin gravity; for the case without matter, alternative expressions 
have been proposed in Refs.~\cite{Ammon&al,deBoer&Jottar,Castro&Llabres}.
Given, however, that to the leading order in the mass parameter the higher-spin fields
do not affect the action, one may wonder if to this order the RT formula remains 
applicable as well. Here, we compare results obtained from that formula 
to those from a direct lattice computation in field theory.

For a single interval, application of the RT formula amounts to computing 
the length of the geodesic, $x(r)$, connecting two given points $x = \pm x_m$ on the boundary.
As before, the boundary is at a large $r = r_m$. The length of the interval is
\be
l = 2 x_m \, .
\label{lint}
\ee 
The geodesic has two symmetric branches: one, with $x < 0$, 
going from $r =r_m$ to the tip at $r = \rtip$, and the other, with $x > 0$, from
$\rtip$ back to $r_m$. The length of the geodesic in the metric (\ref{ds2}) is given by 
\be
\cA[x(r),F(r),G(r)] = 2 \int_{\rtip}^{r_m} [G(r) + F(r) (x')^2]^{1/2} dr \, ,
\label{length}
\ee
where $x' \equiv dx/dr$.
The geodesic equation for $x(r)$ is obtained by extremizing this at fixed $r_m$ and
$x_m$, with $\rtip$ obtained as a function of these parameters in the course of the
procedure. The RT formula for the EE \cite{Ryu&Takayanagi}, $S_{ent} = \cA / (4 G_N)$, 
combined with the relation \cite{Brown&Henneaux} $G_N = 3 / (2 N)$, then gives
\be
\Sent = \frac{1}{6} N \cA \, .
\label{Sent}
\ee

A simplification specific to $d=2$ is that the integrand in \eq{length} is 
independent of $x$, so the corresponding canonical ``momentum'' is conserved along the 
geodesic:
\be
\frac{F(r) x'}{[G(r) + F(r) (x')^2]^{1/2}} = \mbox{const} = [F(\rtip)]^{1/2} \, .
\label{const}
\ee
The value of the constant has been found by noting that $x' \to \infty$ at the tip. 
Moreover, the Hamilton-Jacobi theory applied to the functional (\ref{length}) tells us
that the same constant appears as the derivative of $\cA$ with respect to the endpoint:
\be
\dpar{\cA}{x_m} = 2 [F(\rtip)]^{1/2} \, .
\label{dAdxm}
\ee
With the help of \eqs{lint}{Sent}, this can be expressed as the derivative 
of the EE with respect to the length of the interval:
\be
\dpar{\Sent}{l} = \frac{N}{6} [F(\rtip)]^{1/2} \, .
\label{dSdl}
\ee

Expression (\ref{dSdl}) is curious in its own right, but is not the most convenient one
if we are looking specifically at the case of small deformations
(by a scalar of small amplitude $A$): 
it requires us to consider variations of both $F(r)$ and $\rtip$, i.e., of the 
geodesic itself. In this case, we have found it more convenient to proceed directly from
\eq{length}. Then, to the leading order in the scalar amplitude, we can use the geodesic 
that was the extremum of $\cA$ for the undeformed BTZ background (\ref{btz}):
\be
x_0(r) = \frac{1}{\alpha} \ln 
\frac{c r + \alpha (r^2 - \rtip^2)^{1/2}}{\rtip (r^2 + \alpha^2)^{1/2}} \, ,
\label{xr}
\ee
where $c \equiv (\alpha^2 + \rtip^2)^{1/2}$ is the value of the constant (\ref{const}) for
the undeformed case. Note that, here,
$\rtip$ is the value of $r$ at the tip of the undeformed geodesic (\ref{xr}). As such, 
it is different 
(by terms of order $A^2$) from $\rtip$ we would have to use in \eq{dSdl}. 
On the other hand, in the method employed in what follows, to the required accuracy 
the two values are interchangeable, so we do not use separate notation for each.

For numerical work, it is convenient to 
choose $\rtip$ at will and map it by \eq{xr}  
to a value of $l = 2 x_m$. For future use, we present here a table of $(\rtip, l)$ pairs
obtained in this way (Table \ref{tab:map}).

\begin{table}
\begin{center}
\begin{tabular}{|c|ccc|}
\hline
$\rtip / \alpha$ & 0.5 & 1 & 2 \\
\hline
$l \alpha$ & 2.887 & 1.763 & 0.962 \\
\hline
\end{tabular}
\end{center}
\caption{\small The length $l$ of the entangling interval 
(in units of the thermal wavelength $\alpha^{-1}$, where $\alpha = 2 \pi T$) 
as given by the undeformed geodesic 
(\ref{xr}) for cutoff radius $r_m / \alpha =10^3$ and various values of $\rtip$.}
\label{tab:map}
\end{table}

The metric corrections are found from the $xx$ and $rr$ components of the 
Einstein equations,
\ba
\frac{G'}{G} + 2 G r & = & 2 \kappa^2 r \left[ 
- \frac{q^2 G}{F} \phi^2  + (\phi')^2 + M^2 G \phi^2 \right]  \, , \label{Exx} \\
\frac{F'}{F} - 2 G r & = & 2 \kappa^2 r \left[ 
- \frac{q^2 G}{F} \phi^2  + (\phi')^2 - M^2 G \phi^2 \right]  \, . \label{Err} 
\ea
We now use $\phi$ to denote the radial dependence of the scalar and assume it real.
The original scalar field is now $e^{iqx} \phi(r)$. 
Primes denote derivatives with respect to $r$. These equations have to be solved with
the boundary conditions
\be
G(0) = 1 / \alpha^2 \, ,  \hspace{3em} F(r_m) = r_m^2 \, , \label{bcGF} 
\ee
which correspond to varying the metric with the temperature and length of the $x$ circle
fixed. The boundary
conditions for $\phi$ can be found from \eq{phi} (with $\lam = 1$), which is applicable 
since we are working in the linearized theory. We have
\be
\phi(0) = \tA / \alpha \, , \hspace{3em} \phi'(0) = 0 \, ,
\label{bcphi}
\ee
where $\tA$ is related to $A$ by \eq{tA}. The system consisting
of \eqs{Exx}{Err} and the equation for
the scalar can now be solved numerically, and the solution can be used together with 
the geodesic (\ref{xr}) in \eq{length}. 

\begin{figure}
\begin{center}
\includegraphics[width=4in]{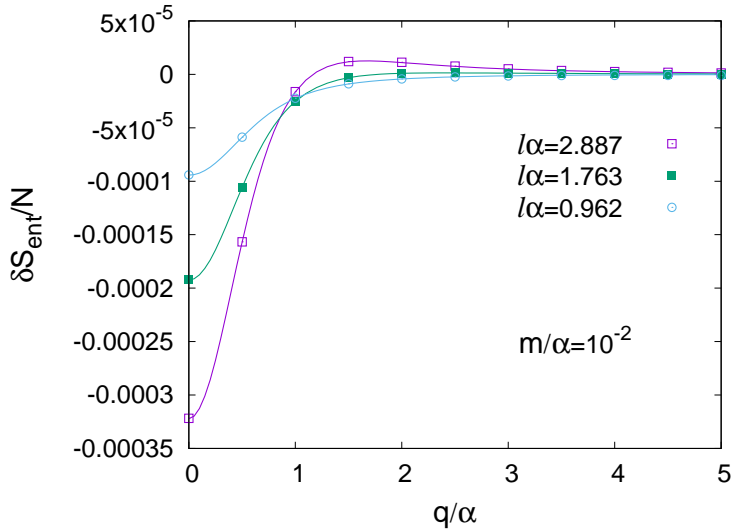}
\end{center}                        
\caption{\small Entanglement entropy change per fermion due to a
mass deformation for intervals of
different lengths, as a function of the superflow
momentum $q$ (in units of $\alpha = 2\pi T$). Curves: results of a holographic 
calculation using general relativity as the dual. 
Points: results of a direct calculation in 
lattice field theory. The interval lengths are the same as in Table~\ref{tab:map}. 
}
\label{fig:EE}
\end{figure}

In Fig.~\ref{fig:EE}, we plot results obtained by this method for 
$m / \alpha = 10^{-2}$, $r_m / \alpha = 10^3$, 
and the values of $\rtip$ shown in Table~\ref{tab:map}.
The length of the geodesic (\ref{xr}) in the undeformed geometry is
\be
\cA_0 = 2 \ln \left( y + \sqrt{y^2 - 1} \right) \, ,
\label{cA}
\ee
where $y \equiv r_m / \rtip$, and we plot the quantity
\be
\frac{\delta \Sent}{N} = \frac{1}{6} (\cA - \cA_0) \, ,
\label{dSent}
\ee
which is the entropy change per fermion, due to a finite mass. This quantity
goes to a finite limit at large $r_m$.

For comparison, in the same figure, we show results of a direct lattice computation 
of the EE for the corresponding values of the interval length $l$ (Table~\ref{tab:map}).
The method is described (for $q=0$) in Ref.~\cite{Casini&al}. We use staggered fermions
\cite{Susskind} with antiperiodic boundary conditions on a uniform lattice with an even 
number $N$ of lattice sites. The Hamiltonian corresponding
to the additional Lagrangian of \eq{Ladd} is discretized as
\be
H_{add} = \frac{q}{2} \sum_{n=0}^{N-1} \Psi_n^\dagger \Psi_n \, ,
\label{Hadd}
\ee
where $\Psi_n$ is a one-component Fermi operator. The results in Fig.~\ref{fig:EE} are
for $N = 5\times 10^4$. The good agreement seen in the figure confirms applicability of
general relativity to computation of $O(m^2)$ terms in the EE.

Let us comment on short-distance cutoffs required in both calculations.
On the gravity side, the cutoff is represented by the maximum radius $r_m$ 
\cite{Susskind&Witten}; on the lattice, by the lattice spacing $h$.
Because the difference (\ref{dSent}) is finite in the limit $r_m \to \infty$, the precise 
relation between $r_m$ and $h$ does not matter, as long as $h$ remains much smaller than
all the physical length scales. For definiteness, we have set $h = 1/r_m$. 
Then, for instance, for $r_m / \alpha =10^3$  
(the value used for compiling Table~\ref{tab:map}),
$1 / (h \alpha) = 10^3$, so $l$ in units of the lattice spacing is obtained by multiplying 
an entry for $l \alpha$ in Table~\ref{tab:map} by 1000.

\section{Limit of a short interval}
\label{sec:short}

In the limit of a short interval, $l \ll \xi$, the leading terms in 
 $\delta S_{ent}$ can be computed analytically. Indeed, in this case, the entire
geodesic lies near the boundary, and we can 
use the asymptotic expression (\ref{phiasymp}) for
$\phi$ and linearized Einstein equations to find the leading logarithms 
in the metric functions there.
The computation is similar to those done in Refs.~\cite{Liu&Mezei,Klebanov&al,Nishioka} 
for the vacuum AdS deformed by scalars
of different masses. In our case, we need to keep track of deformations of both $G$ and $F$,
since both contribute to the geodesic length (\ref{length}). Define the deformations
$G_1$ and $F_1$ by
\be
G(r) = G_0(r) + G_1(r) \, , \hspace{3em} F(r) = F_0(r) + F_1(r) \, .
\label{FG1}
\ee
In the asymptotic region, linearized \eqs{Exx}{Err} become
\ba
\frac{G_1'}{G_0} + 4 r G_1 & = & 
\frac{2 \kappa^2 A^2}{r^3} \left[ - 2 \ln( r\xi) + 1 + \dots \right] , \label{linG} \\ 
\frac{F_1'}{F_0} - \frac{2 r F_1}{F_0^2} & = & 
2 r G_1  + \frac{2 \kappa^2 A^2}{r^3} \left[ 2 \ln^2(r\xi) - 2 \ln (r\xi) + 1 + \dots \right] .
\label{linF}
\ea 
Here and in the next equation dots denote terms suppressed by inverse powers of $r$. 
The solution to the first of these is
\be
G_1(r) = \frac{2 \kappa^2 A^2}{r^4} 
\left[ - \ln^2( r\xi) + \ln r + \mbox{const}  + \dots \right]
\label{G1}
\ee 
Substituting this into the equation for $F_1$, we find that logarithmic terms on the
right-hand side all cancel,
and the leading behavior of $F_1$ at large $r$ is a constant.

To the linear order in $F_1$ and $G_1$, the change in the length 
(\ref{length}) is
\be
\delta \cA[x(r),F(r),G(r)] =\int_{\rtip}^{r_m} 
\frac{G_1(r) + F_1(r) (x')^2}{[G_0(r) + F_0(r) (x')^2]^{1/2}} dr \, .
\label{dcA}
\ee
As before, we can use here the undeformed geodesic (\ref{xr}) since it was extremal to the
zeroth order. The term with $F_1$ does not produce any logarithms of $\rtip$, so as far
as those are concerned
\be
\delta \cA[x(r),F(r),G(r)] \approx \int_{\rtip}^{r_m} (r^2 - \rtip^2)^{1/2} G_1(r) dr \, .
\label{dcA2}
\ee
The integral is convergent in the limit $r_m \to \infty$, and we assume this limit in
what follows. Substituting \eq{G1} for $G_1$ and computing the integrals, we obtain
\be
\delta \cA[x(r),F(r),G(r)] = \frac{2 \kappa^2 A^2}{3 \rtip^2}
\left[ - \ln^2(\rtip \xi) + \left(2 \ln 2 - \frac{5}{3} \right) \ln \rtip + O(1) \right] \, ,
\label{dcA3}
\ee
where $O(1)$ refers to counting of powers of $\ln \rtip$. 
For a near-boundary geodesic, \eq{xr} can be approximated by
\be
x(r) = \left( \frac{1}{\rtip^2} - \frac{1}{r^2} \right)^{1/2} \, .
\label{xapprox}
\ee
In the limit $r_m \to \infty$, this gives $l = 2 x_m = 2 / \rtip$. Finally, using
using \eq{A} to relate $A$ to the fermion mass, we obtain 
\be
\delta \Sent = \frac{1}{6} N m^2 l^2 \left[ - \ln^2(l / \xi) + \frac{5}{3} \ln l +
O(1) \right] \, ,
\label{dSln}
\ee
where we now count powers of $\ln l$.
The leading dependence of \eq{dSent} on $q$ is due to $\ln \xi$ in
\be
\ln^2 (l / \xi) = \ln^2 l -  2 \ln l \ln \xi + O(1) \, 
\label{ln2}
\ee
with $\ln \xi$ given by \eq{lnxi}. 
Note that, in the limit $q \gg T$, expression (\ref{ln2}) becomes
independent of the temperature.

Our results so far have been for the case $m \ll 1/\xi$, where $\xi$ is determined by the
larger of $T$ and $q$. Let us comment on the counterpart of \eq{dSln} for the opposite case,
when $m$ is the largest mass scale in the problem. For example, let us consider deformations 
of the global AdS$_3$ by a scalar of the form (\ref{phi}) subject to the condition
$|q| \ll m$. This is appropriate for the system at zero temperature and a 
low superflow speed. The metric is 
\be
ds^2_{T=0} = F(r) d\tau^2 + r^2 dx^2 + G(r) dr^2 \, .
\label{ads}
\ee
The undeformed AdS of unit radius corresponds to $F(r) = [G(r)]^{-1} = 1 + r^2$. The dual CFT 
now lives on a unit circle: $L_x = 2\pi$. As before, we focus on a scalar of 
mass $M^2 = -1$ corresponding to the operator $O = \bPsi_R \Psi_L$ in the free-fermion CFT.
The amplitude $A$ is related to the mass $m$ of the fermion by the same \eq{A}.
The key difference from the preceding case is 
that the correlation length in the CFT is now determined by the mass,
and the role of the thermodynamic limit is taken over by the condition
\be
m \gg 2\pi / L_x = 1 \, .
\label{mcond}
\ee 
For this condition to apply, the 
amplitude of the scalar must be relatively large, and as result 
the metric deformation cannot be 
small everywhere. Indeed, we can define the correlation length holographically as the radius
$r$ of the region where the metric correction is comparable to the unperturbed metric. 

As the scalar decreases at large $r$, eventually, at $r \gg 1 / \xi$, the metric deformation 
becomes small. We are back in the domain of linearized theory, where the asymptotic formula 
(\ref{phiasymp}) for the scalar applies. The correlation length $\xi$ in it, 
however, is now determined primarily by the mass, and the linearized theory provides no
information on it beyond the estimate $\xi \sim m^{-1}$. Working 
in parallel with the computation above, we obtain the same \eq{dSln} for short intervals;
however, uncertainty in the estimate $\xi \sim m^{-1}$ prevents one from determining 
the coefficient of the $\ln l$ term. The result is 
\be
\delta \Sent = \frac{1}{6} N m^2 l^2 \left[ - \ln^2(ml) + O(\ln l) \right] \, ,
\label{dSm}
\ee
for $ml \ll 1$. We note that the leading correction
in this case coincides with that obtained
in Ref.~\cite{Casini&al} by a direct calculation in field theory. Agreement
up to a numerical factor, between the $\ln^2 l$ terms in the holographic and field-theory 
calculations for $\D = 1$, has been noted in Ref.~\cite{Nishioka}. Here, we show that 
the coefficients match as well.

\section{Discussion}
\label{sec:disc}
In this paper, we have aimed to understand conditions under which the entropies of the
$d=2$ free-fermion CFT deformed by an $x$-dependent mass term $m e^{iqx}$ 
can be computed holographically using standard general relativity as the dual. 
We have considered both the thermal entropy (of the entire space) and 
the entanglement entropy (EE) of an interval. In both instances, we have found that
one can use general relativity to compute the entropy to order $m^2 \xi^2$, where
$\xi$ is the correlation length set by the larger of the temperature and the momentum $q$.
For the thermal entropy, this can be seen as a consequence of the proposed duality
\cite{Gaberdiel&Gopakumar} between this CFT and the higher-spin theory of 
Ref.~\cite{Prokushkin&Vasiliev}: to order $m^2$, the effective actions in that theory
and in general relativity are the same.

For the EE, the reasoning is less direct since in that case, before applying duality, 
one makes transformations of the CFT coordinates and metric
\cite{Casini&al:Derivation,Lewkowycz&Maldacena}. One could argue, however, that, while
these transformations can make the coordinate dependence of the fermion mass complicated,
they do not affect counting of the powers of $m$, so to compute the entropy to order $m^2$
one may still be able to replace the higher-spin gravity dual with the standard one. 
Our results lend support to this argument. 

Finally, we remark that holographic formulas for the EE in higher-spin AdS$_3$ gravity 
without matter have been proposed (within the Chern-Simons formulation) in 
Refs.~\cite{Ammon&al,deBoer&Jottar,Castro&Llabres}. One may expect that, in the presence of 
a scalar necessary to describe a mass deformation, these formulas will have to be modified.
It would be interesting to find out how.

\section*{Acknowledgments}

We thank M. Kruczenski for discussions. 
This work was supported in part by the DOE QuantISED program (through Grant No. DE-SC0019202 
and the consortium ``Intersections of QIS and Theoretical Particle Physics'' at Fermilab) 
and by the W. M. Keck Foundation.

\end{document}